\documentclass[conference]{IEEEtran}
\IEEEoverridecommandlockouts
\usepackage{cite}
\usepackage{amsmath,amssymb,amsfonts}
\usepackage{algorithmic}
\usepackage[ruled,vlined,linesnumbered]{algorithm2e}
\SetKwInput{KwInput}{Input}
\SetKwInput{KwOutput}{Output}
\usepackage{graphicx}
\usepackage{geometry}
\usepackage{comment}
\newcommand{\CLASSINPUTtoptextmargin}{2.4cm}
\newcommand{\CLASSINPUTbottomtextmargin}{4.6cm}
\usepackage[singlespacing]{setspace} 
\usepackage{subfigure}
\usepackage{textcomp}
\usepackage{xcolor}
\def\BibTeX{{\rm B\kern-.05em{\sc i\kern-.025em b}\kern-.08em
    T\kern-.1667em\lower.7ex\hbox{E}\kern-.125emX}}
\begin{document}

\title{Statistical Analysis of the Properties of Geometric Network with Node Mobility
}

\author{\IEEEauthorblockN{1\textsuperscript{st} Md. Arquam}
\IEEEauthorblockA{\textit{\small{Dept. Computer Science and Engineering}} \\
\textit{\small{IIIT Sonepat}}\\
\small{Haryana, India} \\
md.arquam@iiitsonepat.ac.in}
\and
\IEEEauthorblockN{2\textsuperscript{nd} Utkarsh Tiwari}
\IEEEauthorblockA{\textit{\small{Dept. Computer Science and Engineering}} \\
\textit{\small{Shiv Nadar Institute of Eminence}}\\
\small{Delhi-NCR, India} \\
ut353@snu.edu.in}
\and
\IEEEauthorblockN{3\textsuperscript{rd} Suchi Kumari}
\IEEEauthorblockA{\textit{\small{Dept. Computer Science and Engineering}} \\
\textit{\small{Shiv Nadar Institute of Eminence}}\\
\small{Delhi-NCR, India} \\
suchi.singh24@gmail.com}
}

\maketitle

\begin{abstract}
    The movement changes the underlying spatial representation of the participated mobile objects or nodes. In real world scenario, such mobile nodes can be part of any biological network, transportation network, social network, human interaction, etc. The change in the geometry leads to the change in various desirable properties of real-world networks especially in human interaction networks.  In real life, human movement is concerned for better lifestyle where they form their new connections due to the geographical changes.  Therefore, in this paper, we design a model for geometric networks with mobile nodes (GNMN) and conduct a comprehensive statistical analysis of their properties. We analyze the effect of node mobility by evaluating key network metrics such as connectivity, node degree distribution, second hop neighbors, and centrality measures. Through extensive simulations, we observe significant variations in the behavior of geometric networks with mobile nodes.
\end{abstract}

\begin{IEEEkeywords}
Complex Network, Epidemic Spreading, Contact Network, Random Geometry, Connecting Distance 
\end{IEEEkeywords}

\section{Introduction}\label{sec:Introduction} 
Over the past two decades, networks have become essential for constructing models that replicate the structural properties of numerous real-world systems. Various network models have been proposed to measure the significance of structural properties on/of network dynamics. Some well-known models include the Erdős-Rényi model \cite{estrada2012structure}, small-world networks \cite{barrat2000properties}, and scale-free networks \cite{barabasi2003scale}. One variant of random networks, random geometric graph (RGG) models, incorporates the spatial distribution of nodes in a network. These models have recently gained attention as an alternative to simpler yet less realistic models such as the ubiquitous Erds-Rényi model \cite{estrada2012structure}. Geometry plays a crucial role in understanding the structure and properties of real-world networks. Due to the geometric structure of RGGs, this model is particularly appealing for applications in wireless network modeling \cite{jia2004wireless}, social networks \cite{hoff2002latent}, and biological networks \cite{higham2008fitting}. In many of these real-world networks, the probability of a connection between two agents (nodes) depends on the distance between their positions in a metric space \cite{penrose2003random}.

Since the geometry of the nodes changes with their movement in the network, mobility pattern analysis has become a crucial research domain associated with RGG models. Initially, RGGs were used for wireless networks, but nowadays, they are widely applied in the analysis of disease spread and rumor dissemination. Some researchers have studied the interconnection between human mobility and epidemic spreading within contact networks formed by individuals who come into contact with each other. In all these scenarios, analyzing human mobility patterns provides a more realistic and accurate understanding of the performance of mobile network protocols.
Commonly used mobility models include random waypoint analysis \cite{bettstetter2003node}, random walk approaches such as Brownian motion \cite{yu2018identifying}, and Markovian mobility \cite{akyildiz2000new}. However, these models cannot guarantee an accurate analysis of nodes' mobility. Levy \textit{et al.} \cite{rhee2011levy} compared human walking patterns to Levy walks, which follow a heavy-tailed flight and pause distribution with super-diffusive mobility. Duchemin \textit{et al.} \cite{duchemin2023random} surveyed the applications of RGGs in high-dimensional settings, non-parametric inference, and discussed multiple variants of RGGs. They observed geometric properties in both dense and sparse regimes and evaluated nonparametric inference within the context of RGGs. In addition, some growth models were included and the connection between RGGs and community-based models was explored.
 
In the classical Random Geometric Graph (RGG) model, introduced by Gilbert \cite{penrose2003random}, $n$ independent and identically distributed geometrical points $\{X_i\}_i \in R^d$ are considered. The graph is constructed with the vertex set $N=\{1,2,...,n\}$, where nodes $i$ and $j$ are connected if and only if the Euclidean distance $\parallel X_i - X_j \parallel_d$ is smaller than a predefined threshold called connecting radius $r$. In other words, two points are connected only if their distance is smaller than $r$. Thus, the randomness in this model arises from the sampling of the geometric points according to a certain distribution \cite{penrose2003random}. Arquam \textit{et al.} \cite{arquam2020epidemic, arquam2019modelling} extended the RGG model to analyze epidemic spreading on human contact networks, considering changes in the geometric location and connecting distance of two individuals, but failed to explain the impact of the connecting distance. 

In all the papers reviewed in the literature survey, mobility, velocity, and the resting time of the nodes in the network have not been considered. If a node's resting time is longer, it will form stronger connections with its neighboring nodes. Therefore, in this paper, we analyze some important parameters to investigate the statistical properties of Random Geometric Networks with Mobile Nodes (GNMN). Our focus is on analyzing the behavior of degree distribution, variation in centrality, and the second neighbor of GNMN. Degree distribution explains the behavior of network connectivity, while centrality indicates the importance of nodes and edges. The second neighbor describes the propagation pattern in the network, which may be applied to model epidemic spreading and information propagation. Hence, the objectives of this manuscript are as follows:

\begin{itemize}
    \item Generate a random geometric network with mobile nodes.
    \item Analyze network properties such as centrality measures, second-hop neighbors, degree distribution, and disconnected components by varying parameters like radius, rest time, and velocity.
    \item Study the impact of mobility in real-world networks.
\end{itemize}

The paper is organized as follows: Section 1 provides the introduction, recent trends, and applications of GNMN. Section 2 presents a literature survey. Section 3 presents the formulation of key parameter definitions and describes the algorithm analysis of GNMN. Section 4 demonstrates the results and analysis of GNMN. Section 5 discusses the conclusions and future scope of the work.

\section{Literature Survey}

Human movement is influenced by various factors, such as employment opportunities, medical facilities, recreational activities, etc. This movement generally occurs within a predefined geographical region. Human movements are typically modeled using random walk or diffusion-based concepts~\cite{havlin2002diffusion}. Viswanathan et al. proposed a model based on the analysis of monkey and marine predator movements, describing it as Lévy flight~\cite{viswanathan1996levy}. However, this model assumes long-range connections for humans, which is not always accurate. Bettstetter et al. proposed a random waypoint mobility model for wireless ad hoc networks where humans can connect to the nearest sink to transfer data~\cite{bettstetter2003node}. This model allows connections to be established even while humans are moving, which is also unrealistic. Mazalov et al. proposed a model to determine the centrality of nodes and edges using Kirchhoff's law, which helps to identify the importance of nodes in a network~\cite{mazalov2016kirchhoff}. Gonzalez et al. proposed a hybrid model that combines Lévy flight with random walk to describe human movement~\cite{pappalardo2023future}. However, this model requires improvement, as it does not account for connection establishment while moving, which is crucial for understanding epidemic spreading due to local contact.

The stochastic nature of human movement makes it challenging to study large-scale variations in connectivity patterns, clustering, and community formation in real-world systems~\cite{liu2023diffusion, agrawal2020community}. In a human population network, agents' mobility and respective velocities influence epidemic spreading within a given spatial region. Epidemic spreading in human populations can be modeled by considering the random geometry of nodes, where positional changes are non-uniform due to unequal velocities. Humans form connections when they come within each other's connectivity radius. Various models have been proposed to explain epidemic spreading in human connection networks~\cite{guidolin2023innovation,moore2024network}, but most consider these networks to be static.

\section{Proposed Methodology}
In this section, a Random Geometric Network with Mobile Nodes (GNMN) is generated by extending the model proposed in \cite{arquam2020epidemic}. In the proposed model, nodes are randomly distributed in a geometric space, forming connections if they fall within a specified radius $r$. Unlike static nodes, a fraction of the nodes are mobile, changing their positions according to Brownian Motion. These mobile nodes update their connections upon reaching new locations within the specified geometric space, which can be either circular or square. After generating the GNMN, the effects of various network parameters on the GNMN are also studied.

\subsection{Random Geometric Network with Mobile Nodes (GNMN)}

A geometric network is created by following the steps provided in Algorithm \ref{algoRGG}. The process begins by randomly initializing node positions within the given dimensions. Various network parameters can be considered, such as the total number of nodes$N$, the velocity of mobile nodes $v$, the time of rest $t_{rest}$, the time of movement $t_{move}$, the probability of the nodes remaining in a static position $p_{stat}$, the connection radius $r$ and the number of mobile nodes $n_{moves}$. The network is generated through a series of radius values, updating node positions and connections based on node movement and proximity. The position of each node is updated considering the current positions of nodes, the dimensions of the space, the velocity of movement, and the number of nodes to move. A subset of nodes is randomly selected to move, with $n_{moves}$ kept fixed. For each selected node, a new position within the defined dimensions is generated. The movement is checked to ensure it is within a certain distance threshold determined by the velocity, and the position is updated accordingly. Finally, the updated positions and the indices of the moved nodes are provided.

\begin{algorithm}[!htb]
\DontPrintSemicolon
\caption{Random Geometric Network with Mobile Nodes}
\label{algoRGG}
  \KwInput{$N$, velocity, $t_{rest}$, $t_{move}$, dimensions, $pstat$, $r_{value}$, $n_{moves}$}
  \KwOutput{Random Geometric Network $G^{RGN}$}
  $ndim \leftarrow len(dimensions)$ \\
    $positions \leftarrow random.rand(N, ndim) \times np.array(dimensions)$ \\
   $rest_{time} \leftarrow np.empty(N) $ \\
   $prest \leftarrow \frac{t_{rest}}{(t_{rest} + avg(dimensions) \times avg(velocity))}$ \\
   $q0 \leftarrow prest $ \\
   $G \leftarrow nx.Graph()$ \\
   Add nodes to the graph $G$ \\
   $contacts \leftarrow []$ \\
   \For{$i \mbox{ in } range(N)$}
   {$d \leftarrow \sqrt(\sum (positions - positions[i]))^2$ \\
   \If{$d < r_{value} \mbox{ \& } n != i$}
   {$contacts.append(n)$}
   }
   \# Update Nodes position \\
   $old\_positions \leftarrow positions$ \\
   $new\_positions \leftarrow np.empty\_like(positions)$ \\
   $nodes\_to\_move \leftarrow random.choice(len(positions), num\_nodes\_to\_move)$ \\
   \For{$i \mbox{ in } range(len(positions))$}
   {
   \If{$i \mbox{ in } nodes\_to\_move$}
   {
   \While{True}
   {
   $z1 \leftarrow positions[i]$\\
   $z2 \leftarrow random.rand(ndim) \times np.array(dimensions)$ \\
   $d = \frac{\sqrt(\sum ((z2 - z1)^2)}{\sum np.array(dimensions)^2}$ \\
   \If{$random.rand() < d \times velocity$}
   {
   break
   }
   $new\_positions[i] \leftarrow z2$ \\
   }
   }
   \Else{$new\_positions[i] \leftarrow positions[i]$}
   }
   Update the new contact list by repeating steps $9$ - $12$ \\
   return $G^{RGN}$
\end{algorithm}

\subsection{Evaluation Matrics}
In this section, some network parameters is evaluated considering the geometry of the network.  Let $N$ be the total number of nodes, and $A$ be the area of the geometrical region in which the nodes are distributed. Then various network parameters can be evaluated as follows:

\subsubsection*{First Neighbor Probability:}
The probability $P(nei)$ that two nodes are connected and are neighbors depends on their distance $r$ from each other. This probability is proportional to the area of a circle with radius $r$, as described in Eq. \eqref{Eq1}.

\begin{equation}
    P(nei)=\frac{\pi r^2}{A} \label{Eq1}
\end{equation}

Then the expected number of neighbors $E[nei]$ can be evaluated as follows:
$$E[nei]=N*P(nei)=N*\frac{\pi r^2}{A}.$$

\subsubsection*{Second Neighbor Probability:}
The second neighbor of a given node is a node that is within distance $r$ from any of the first neighbors. To find this, we need to consider the probability that a node is within distance $r$ from any node in the neighborhood of the original node. Let's denote the average number of neighbors of a neighbor as $ k_1 $ i.e., average degree of the network), as follows:
    $$ k_1 = E[nei]=N*\frac{\pi r^2}{A}$$
The probability $P(sec\_nei)$ that a node is a second neighbor is the probability that it is connected to any of the $k_1$ first neighbors. This is expressed in Eq. \eqref{Eq2}.   

\begin{equation}
    P(sec\_nei)=1-(1-P(nei))^{k_1} \label{Eq2}
\end{equation}

where $(1-P(nei))^{k_1}$ is the probability that a node is not connected to any of the $k_1$ neighbors.

\subsubsection*{Centrality Measures}
Various centrality measures help to identify the most important nodes in the network. Given that nodes in GNMN are randomly distributed and can be mobile, centrality measures may fluctuate over time. Therefore, some common types of centrality; degree centrality and betweenness centrality, are derived for the geometric network with mobile nodes.
\begin{itemize}

\item  \textbf{Degree centrality} is the simplest form of centrality, defined as the number of direct connections (neighbors) a node has. For a given node $i$, the degree centrality $CD(i)$ can be expressed as,
$$CD(i)=k_i$$ where $k_i$ is the degree of node $i$. The expected degree $k_i$ and degree centrality $CD(i)$ for node $i$ is expressed in Eq. \eqref{Eq3}

\begin{equation}
    E[CD(i)] = E[k_i]= N*P(nei) = N*\frac{\pi r^2}{A} \label{Eq3}
\end{equation}
        
\item \textbf{Betweenness centrality} measures the extent to which a node lies in the shortest paths between other nodes. For node $i$, the betweenness centrality $CB(i)$ is defined as:
$$CB(i) = \sum_{s \neq i \neq t} \frac{\sigma_{st}(i)}{\sigma_{st}}$$
where, $\sigma_{st}$ is the total number of shortest paths from node $s$ to node $t$, $\sigma_{st}(i)$ is the number of those shortest paths that pass through node $i$. In a GNMN, calculating the exact betweenness centrality analytically is challenging due to the randomness and mobility of nodes. Therefore, approximation methods based on the average properties of the network are used. The betweenness centrality in a GNMN can be approximated by considering the expected shortest paths that pass through a given node. Since nodes are uniformly distributed, the probability that a node lies on the shortest path between any two other nodes depends on the node density and the connectivity radius $r$
Let $\rho = \frac{N}{A}$ represents the node density. The average shortest path length $<L>$ in a GNMN can be approximated by Eq. \eqref{Eq4}.

\begin{equation}
    <L> = \frac{\sqrt{A}}{r}. \label{Eq4}
\end{equation}

The expected betweenness centrality $E[CB(i)]$ for node $i$ can be approximated by the fraction of shortest paths it lies on, which is proportional to the area of its connectivity region ($\pi r^2$) relative to the entire network ($A$). This fraction is then multiplied by the total number of shortest paths in the network. The $E[CB(i)]$ can be expressed in Eq. \eqref{Eq5}.
\begin{equation}
    E[CB(i)] \approx \frac{\pi r^2}{A}*\frac{N(N-1)}{2} \label{Eq5}
\end{equation}
where area covered by node $i$ is approximately $\pi r^2$, the total area is $A$, $\frac{N(N-1)}{2}$ represents the total number of shortest paths in the network, assuming the network is dense enough for almost all nodes to be reachable. The approximations provide insights into the expected centrality measures in a dynamically changing geometric network with mobile nodes. As nodes move and their positions change over time, the centrality values can fluctuate accordingly.
\end{itemize}

\section{Results and Analysis}
The random geometric graph is drawn for a network with $N=2000$ nodes. The network is formed considering various parameters such as connecting radius $r$, velocity $v$, time of rest $t_{rest}$, time of movement $t_{move}$, and the total number of nodes selected for the movement $n_{move}$. The detailed score is provided in Table \ref{tab:parameter}.
\begin{table}[!htb]
    \centering
    \begin{tabular}{|c|c|}
        \hline
         Parameter &  Range of values \\ \hline
         $N$ & $2000$ \\
         $v$ & $0.3 - 1.1$ \\
         $r$ & $0.05 - 0.85$ \\
         $t_{rest}$ & $5-25$ \\
         $t_{move}$ & $20$ \\
         $dimension$ & $(12,12)$ \\
         $p_{stat}$ & $0.8$ \\
         $n_{move}$ & $100$\\ \hline
    \end{tabular}
    \caption{Network parameters with the assigned values}
    \label{tab:parameter}
\end{table}

\subsection*{Visualization of GNMN with varying network parameters}
\begin{figure}[!htb]
\begin{center}
$\begin{array}{ccc}
\includegraphics[width=.3\linewidth,height=1.in]{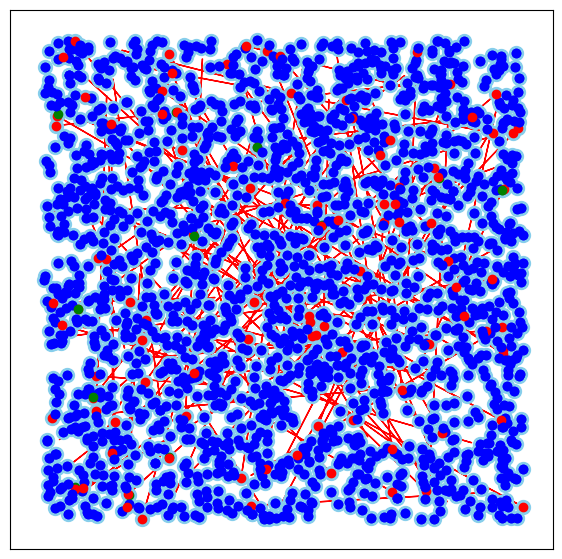}  &
\includegraphics[width=.3\linewidth,height=1.in]{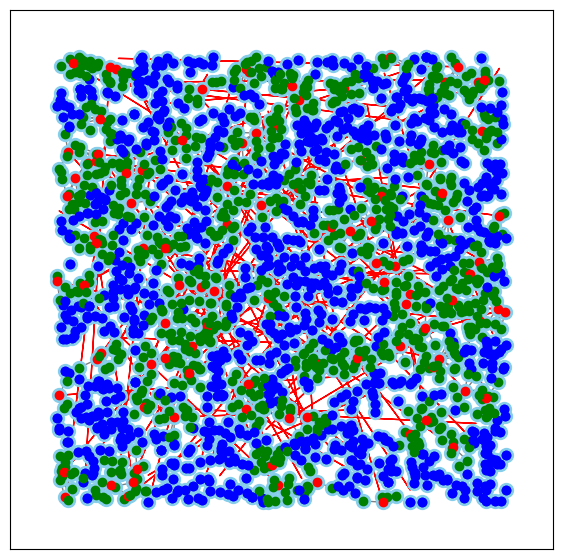}  &
\includegraphics[width=.3\linewidth,height=1.in]{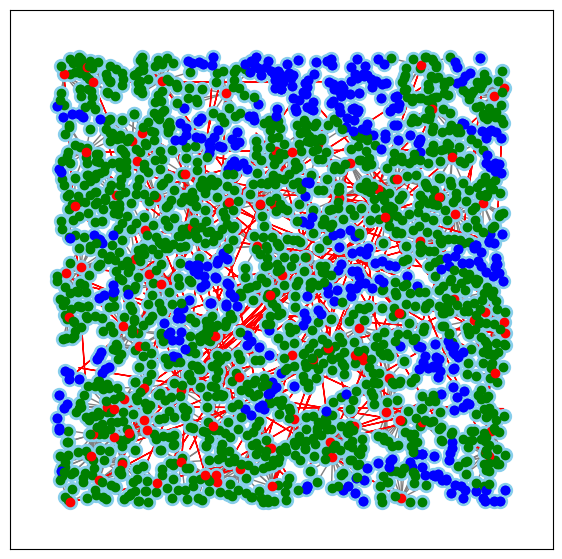} \\
\mbox{(a) $r = 0.05$} & \mbox{(b) $r=0.55$} & \mbox{(c) $r = 0.85$}\\
\includegraphics[width=.3\linewidth,height=1.in]{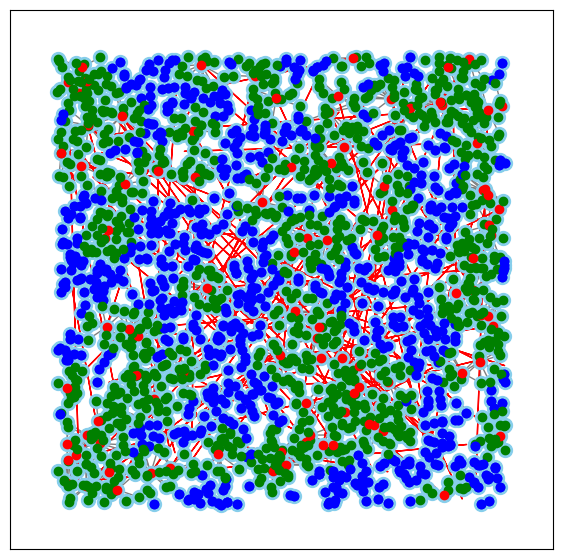}  &
\includegraphics[width=.3\linewidth,height=1.in]{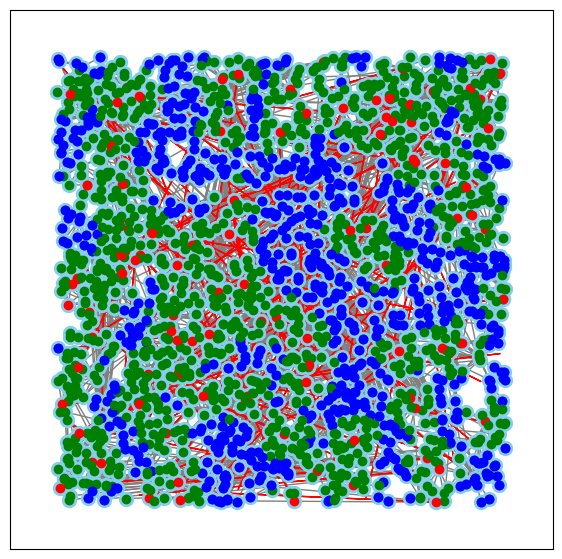}  &
\includegraphics[width=.3\linewidth,height=1.in]{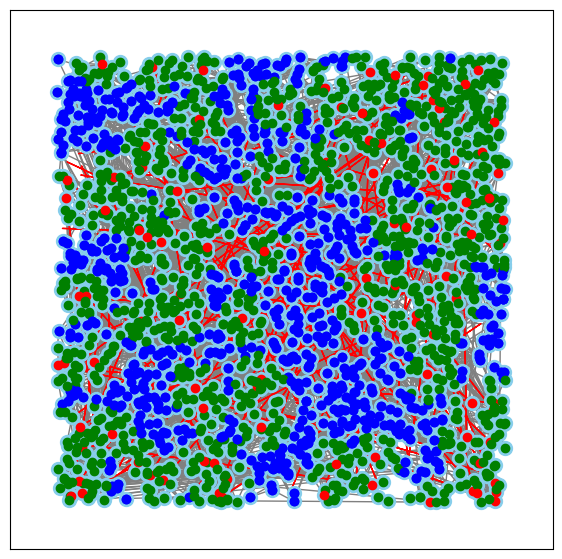} \\
\mbox{(d) $v = 0.3$} & \mbox{(e) $v = 0.7$} & \mbox{(f)$v = 1.1$} \\
\includegraphics[width=.3\linewidth,height=1.in]{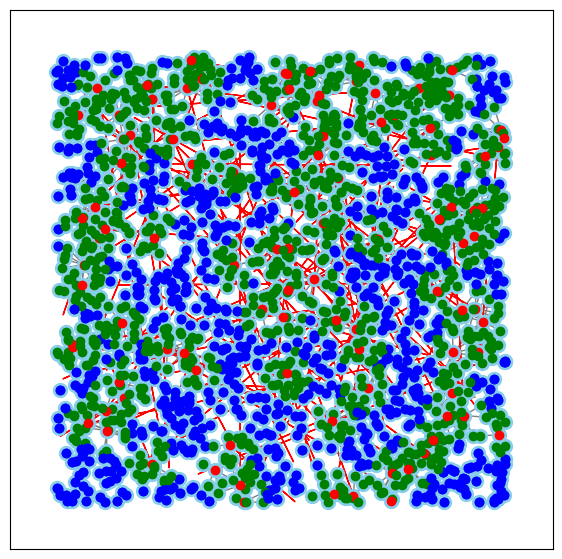}  &
\includegraphics[width=.3\linewidth,height=1.in]{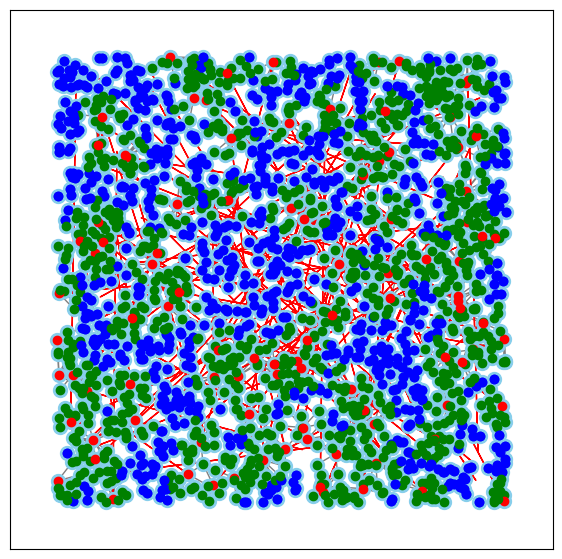}  &
\includegraphics[width=.3\linewidth,height=1.in]{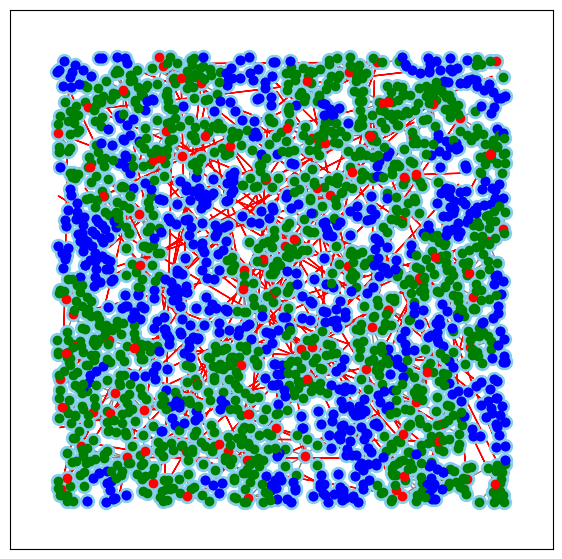} \\
\mbox{(g) $t_{rest} = 5$} & \mbox{(h) $t_{rest} = 15$} & \mbox{(i) $t_{rest} = 25$} \\
\end{array}$
\caption{The random geometric network is formed considering varying radius (a) - (c), velocities (d) - (f), and $t_{rest}$ (g) - (i).}
\label{RGG_varying_r}
\end{center}
\end{figure}

The movement of the nodes is visualized by plotting arrows from old positions to new positions for moved nodes, marking moved nodes in red, marking nodes with new connections in green, and marking disconnected nodes in blue Figure \ref{RGG_varying_r}. The GNMN is visualized by keeping all the parameters constant except $v$, $r$, and $t_{rest}$ (refereed in Table \ref{tab:parameter}). The effect of the radius $r$ on the geometric network is illustrated in Figure \ref{RGG_varying_r} (a) - (c) where $r$ is ranging from $0.05$ to $0.85$, with a constant velocity $v=0.5$, and $t_{rest} = 10$. Initially, when $r = 0.05$, the network is completely disconnected and all the nodes are in blue (disconnected nodes) and red (mobile nodes) in colors (please refer\ref{RGG_varying_r} (a)). With the increasing value of $r$, the probability of having neighbors are increasing (From Eq. \eqref{Eq1}) and the network becomes more connected and the number of green color nodes are increasing (from Figure \ref{RGG_varying_r} (b)-(c)). The analysis of the varying velocity values is presented in Figure \ref{RGG_varying_r} (d) - (f). The velocity ranges from $0.3$ to $1.1$, with a constant radius $r=0.65$, and $t_{rest} = 10$. As the value of $r = 0.65$, the network has multiple connections. But the velocity of the nodes are changing randomly so frequent connections and disconnections take place and it does not effect much on the density of the network. The same happens with the varying values of $t_{rest}$ where the values of $t_{rest}$ range from $5$ to $25$, with a constant radius $r=0.65$ and velocity $v = 0.5$ (presented in Figure \ref{RGG_varying_r} (g) - (i)). Varying $t_{rest}$ also increase the randomness in connectivity. If the value of $t_{rest}$ increases, the number of connections also increases because more nodes come into contact with the node moving that is at rest during their movement.

\subsection*{Second Hop Neighbors}

\begin{figure}[!htb]
   \begin{center}
   $\begin{array}{ccc}
   \includegraphics[width=.3\linewidth,height=1.in]{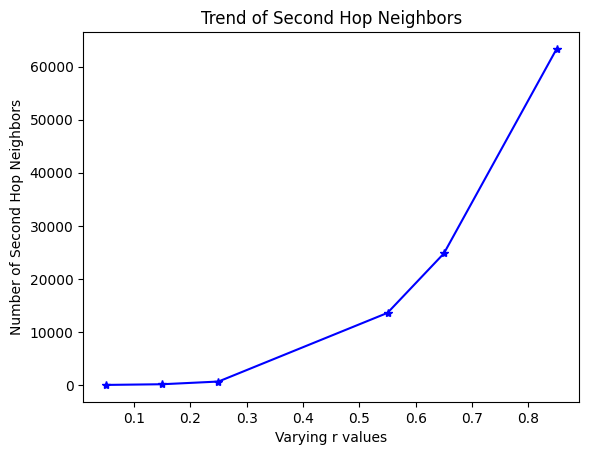}  &
    \includegraphics[width=.3\linewidth,height=1.in]{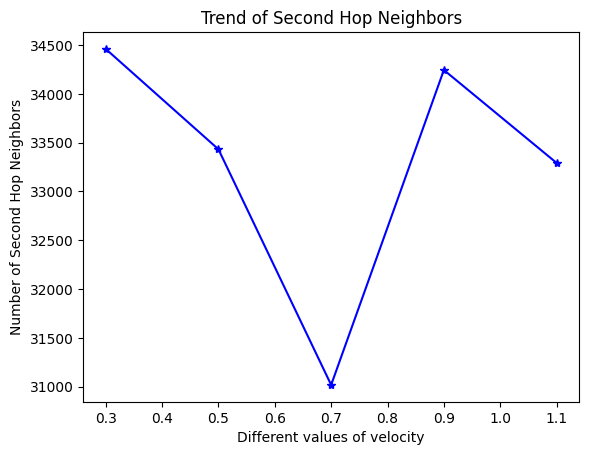} &
     \includegraphics[width=.3\linewidth,height=1.in]{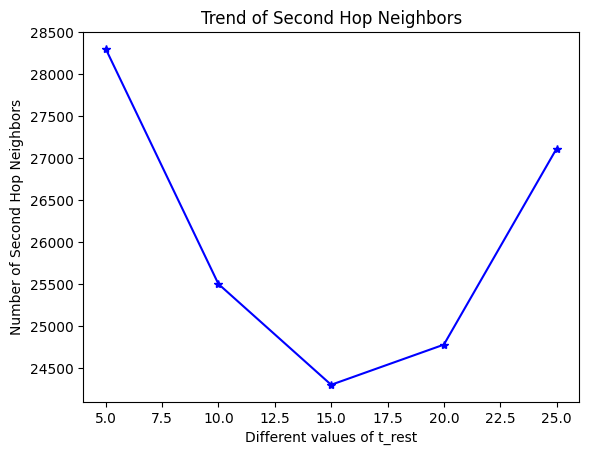} \\
    \mbox{(a)} & \mbox{(b)} & \mbox{(c)} \\
\end{array}$ \caption{Trend of second hop neighbors based on the value of varying radius, velocity and $t_{rest}$}
    \label{fig:secondhop}
    \end{center}
\end{figure}

The $2^{nd}$ Hop neighbourhood provides information about the connectivity and spread of activation to the rest of the network. Therefore, second-hop neighbors pattern of GNMN is plotted in Figure \ref{fig:secondhop} (a-c) considering various parameters. The trend of second-hop neighbors is exponential if connecting radius is increasing. It follows the monotonic function \ref{fig:secondhop} (a). The values for the second hop neighbors and the number of new connections are $(2, 124, 718, 14910, 25780, 64382)$ and $(8, 81, 154, 511, 615, 782)$ for radii $r = 0.05, 0.15, 0.25, 0.55, 0.65, 0.85$, respectively, as shown in Figure \ref{fig:secondhop}(a). While the curve of trend of second-hop neighbors is non-monotonic while considering velocity and $t_{rest}$. Some time, second-hop neighbors are increasing and some time second-hop neighbors is decreasing while increasing the value of velocity and $t_{rest}$ as shown in Figure \ref{fig:secondhop} (b-c).  It increases the new connections and second hop neighbour as shown in Figure \ref{fig:secondhop}(b). The values for the second hop neighbors and the number of new connections are $(26762, 26116, 26242, 27098, 26374)$ and $(609, 592, 584, 633, 596)$ for velocities $v = 0.3, 0.5, 0.7, 0.9, 1.1$, respectively. The values for the second hop neighbors and the number of new connections are $(27456, 25660, 22360, 27584, 26618)$ and $(630, 619, 573, 648, 597)$ for $t_{rest} = 5, 10, 15, 20, 25$, respectively, as shown in Figure \ref{fig:secondhop}(c). 

\subsection*{Degree Distribution}

\begin{figure}[!htb]
\begin{center}
$\begin{array}{ccc}
\includegraphics[width=.3\linewidth,height=1.0in]{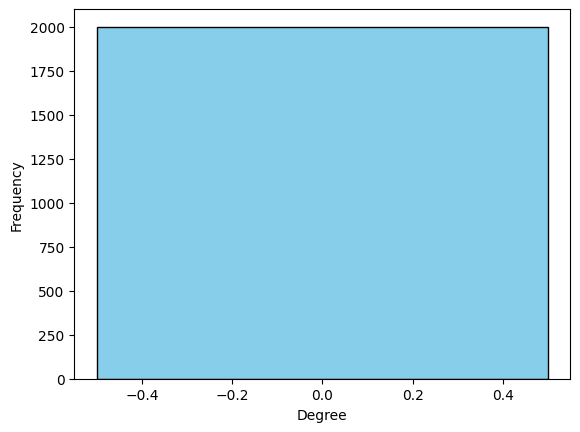}  &
\includegraphics[width=.3\linewidth,height=1.0in]{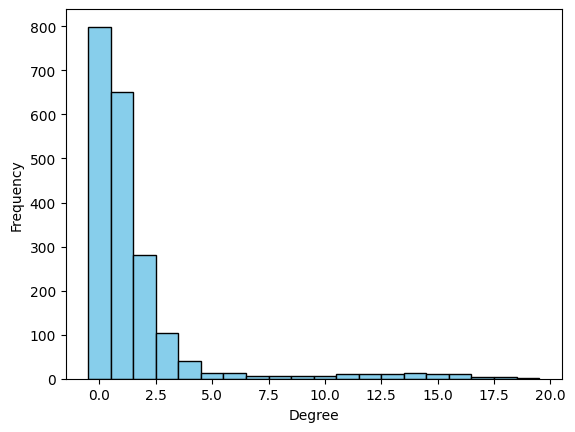}  &
\includegraphics[width=.3\linewidth,height=1.0in]{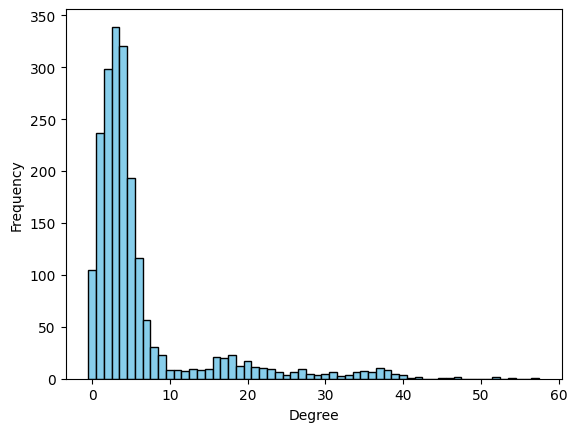} \\
\mbox{(a) $r = 0.05$} & \mbox{(b) $r = 0.55$} & \mbox{(c) $r = 0.85$} \\
\includegraphics[width=0.3\linewidth,height=1.0in]{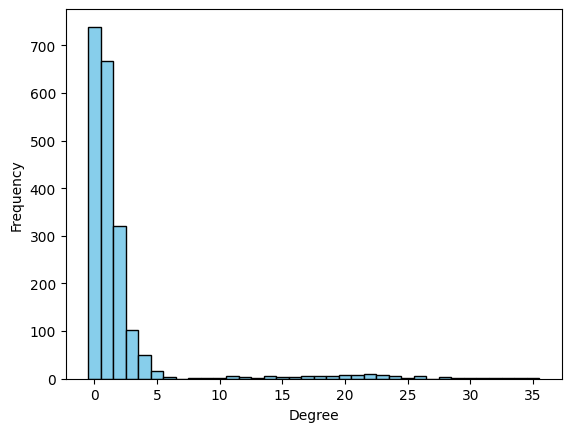}  &
\includegraphics[width=0.3\linewidth,height=1.0in]{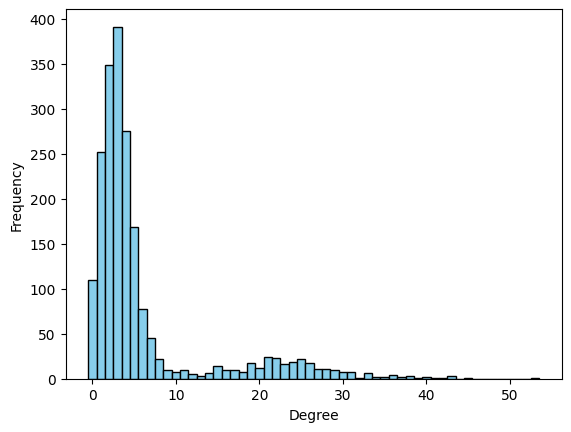}  &
\includegraphics[width=0.3\linewidth,height=1.0in]{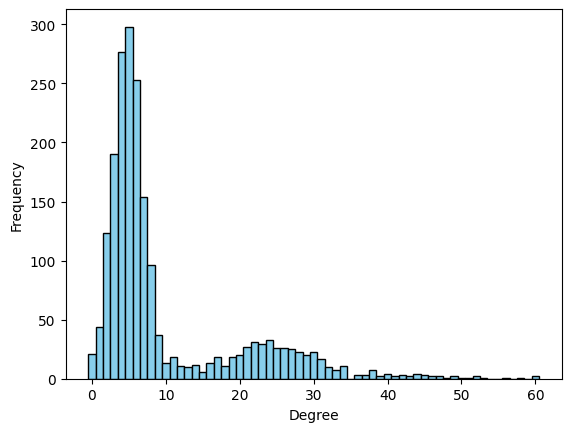} \\
\mbox{(d) $v = 0.3$} & \mbox{(e) $v = 0.7$} & \mbox{(f) $v = 1.1$} \\
\includegraphics[width=0.3\linewidth,height=1.0in]{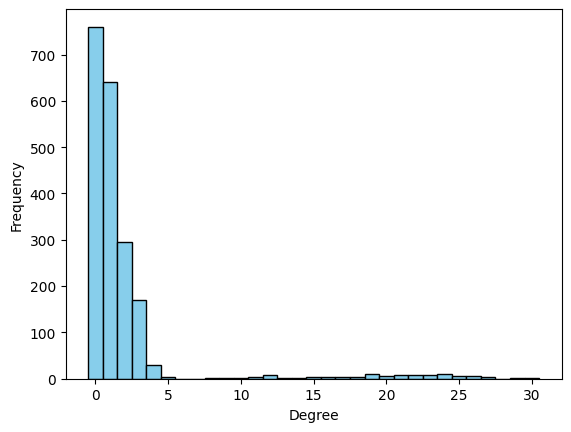}  &
\includegraphics[width=0.3\linewidth,height=1.0in]{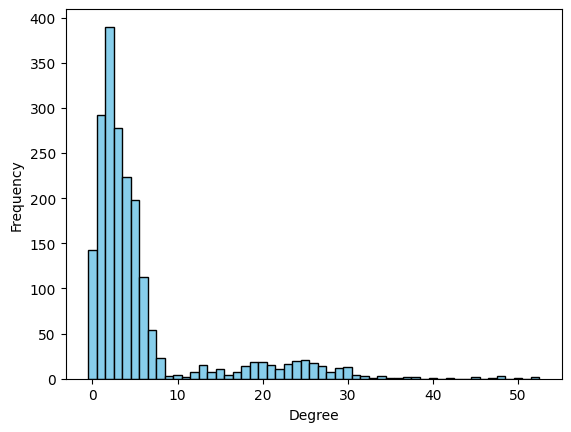}  &
\includegraphics[width=0.3\linewidth,height=1.0in]{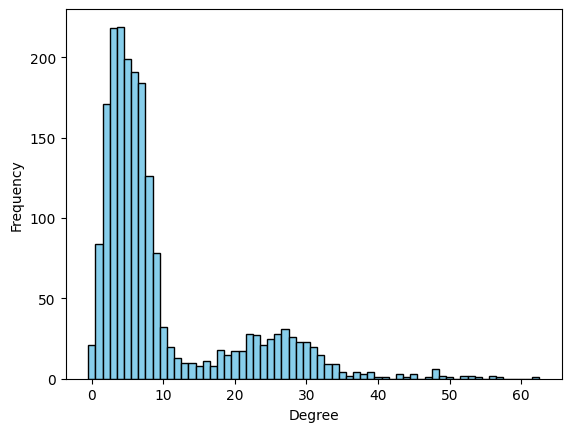} \\
\mbox{(g) $t_{rest} = 5$} & \mbox{(h) $t_{rest} = 15$} & \mbox{(i) $t_{rest} = 25$} \\
\end{array}$
\caption{Degree distribution of the nodes in the random geometric graph plotted for different values of radius, velocity, and $t_{rest}$}
\label{rewire_example}
\end{center}
\end{figure}

The Degree Distribution is plotted against the various parameters as shown in Figure \ref{rewire_example} (a)- (i). The degree distribution looks like a perturbed Poisson distribution.
Increasing the velocity and $t_{rest}$ increase the randomness in connecting pattern of network as it can be seen in Figure \ref{rewire_example} (d - i). Due to randomness, the degree distribution looks like a perturbed Poisson. Increasing the value of connecting radius $r$, increases the number of connections. Due to this degree distribution looks like scale free as shown in Figure \ref{rewire_example} (a - c).

\subsection*{Centrality Analysis}
\begin{figure}[!htb]
\begin{center}
$\begin{array}{ccc}
\includegraphics[width=0.3\linewidth,height=1.0in]{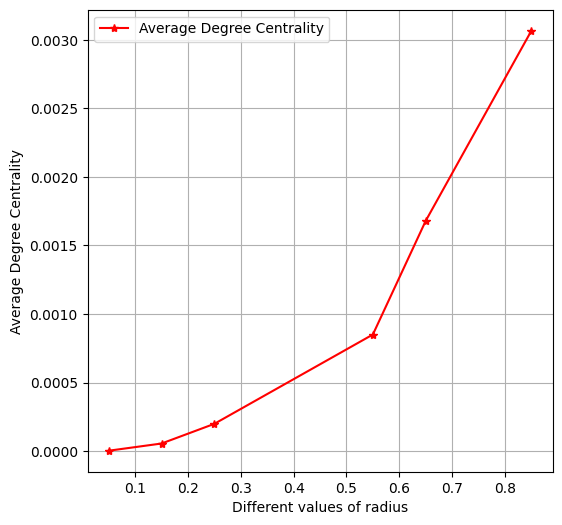}  &
\includegraphics[width=0.3\linewidth,height=1.0in]{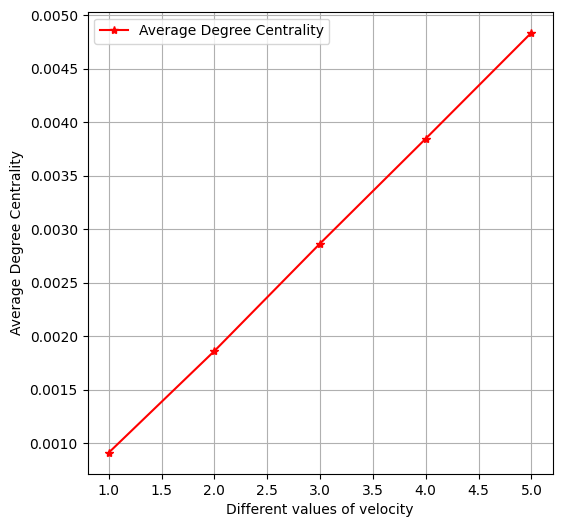} &
\includegraphics[width=0.3\linewidth,height=1.0in]{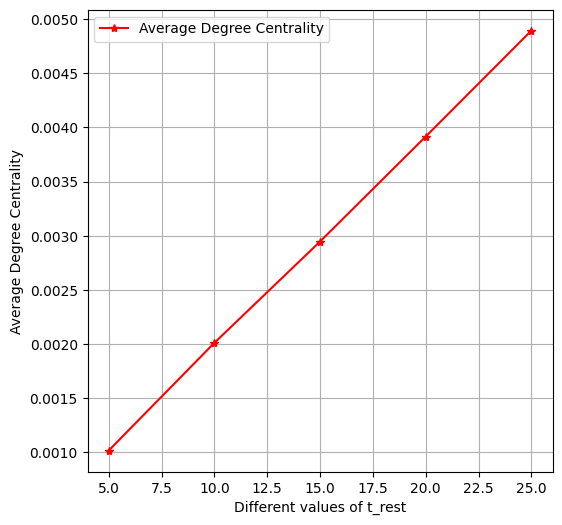} \\
\mbox{(a)} & \mbox{(b)} & \mbox{(c)} \\
\includegraphics[width=0.3\linewidth,height=1.0in]{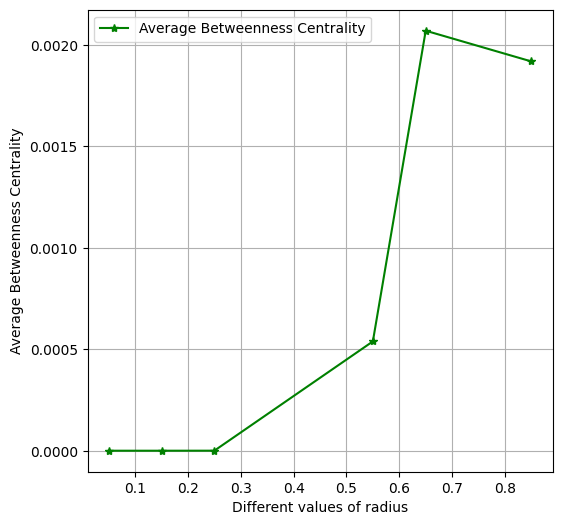}  &
\includegraphics[width=0.3\linewidth,height=1.0in]{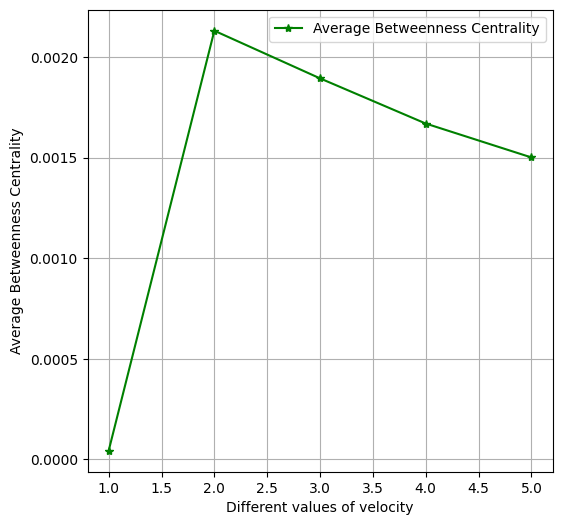}  &
\includegraphics[width=0.3\linewidth,height=1.0in]{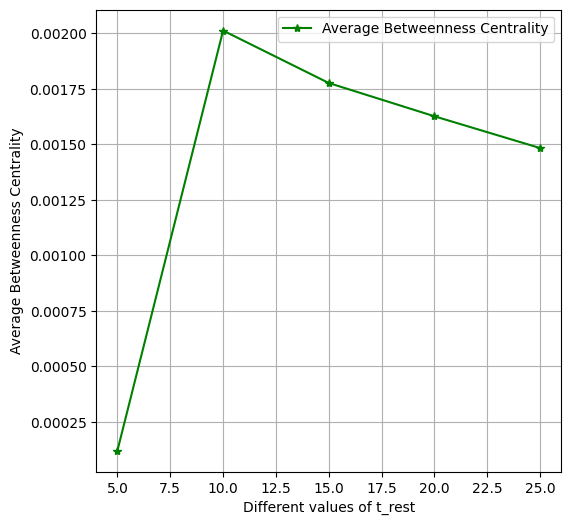} \\
\mbox{(d)} & \mbox{(e)}  & \mbox{(f)} \\

\end{array}$
\caption{Average centrality measures; degree, and betweenness for different values of radius, velocity and $t_{rest}$.}
\label{Centrality_example}
\end{center}
\end{figure}

Centrality measures are an important tool for understanding networks. Therefore, Degree and Betweenness centrality is plotted in Figure \ref{Centrality_example} (a-f) by providing the different values of connecting radius ($r$), Velocity ($v$), and $t_{rest}$. Increasing the value of $r$, degree centrality increases exponentially as shown in Figure \ref{Centrality_example} (a), because, increasing the value of $r$ increase the degree of nodes while increasing the value of $v$ and $t_{rest}$, degree centrality increases linearly as shown in Figure \ref{Centrality_example} ((b) \& (c)), because increasing the value of $v$ and $t_{rest}$ increases the randomness. In case of betweenness centrality, the value of centrality is sometimes increasing and sometimes decreasing while value of $r$,$v$, and $t_{rest}$ are increasing as shown in Figure \ref{Centrality_example} ((d), (e) \& (f)) because betweenness centrality depends on the importance of nodes not on the degree.

\subsection*{Connected Components}

\begin{figure}[!htb]
\begin{center}
$\begin{array}{ccc}
\includegraphics[width=0.3\linewidth,height=1.in]{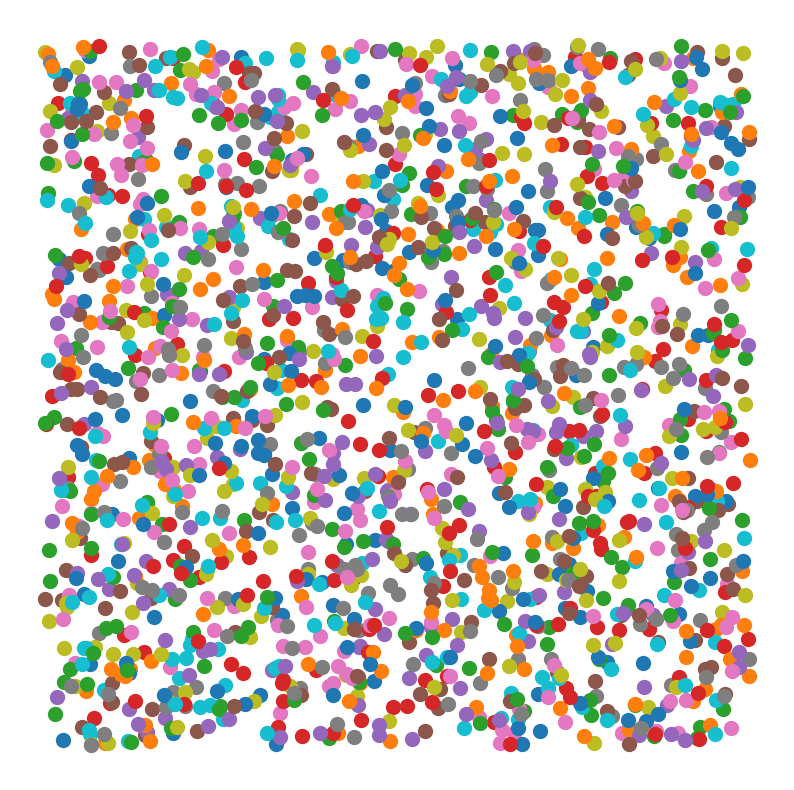}  &
\includegraphics[width=0.3\linewidth,height=1.in]{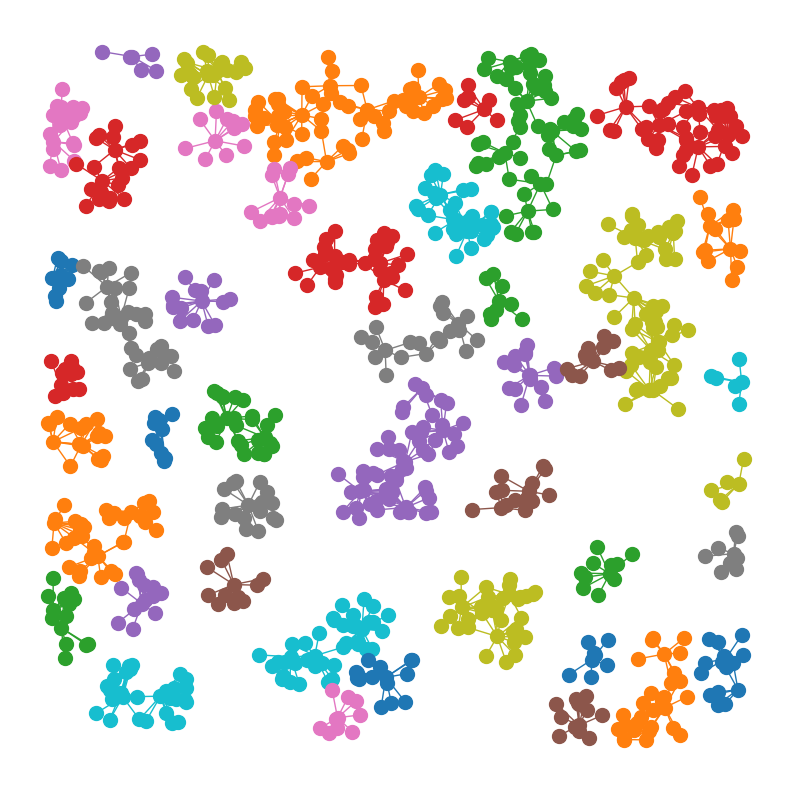}  &
\includegraphics[width=0.3\linewidth,height=1.in]{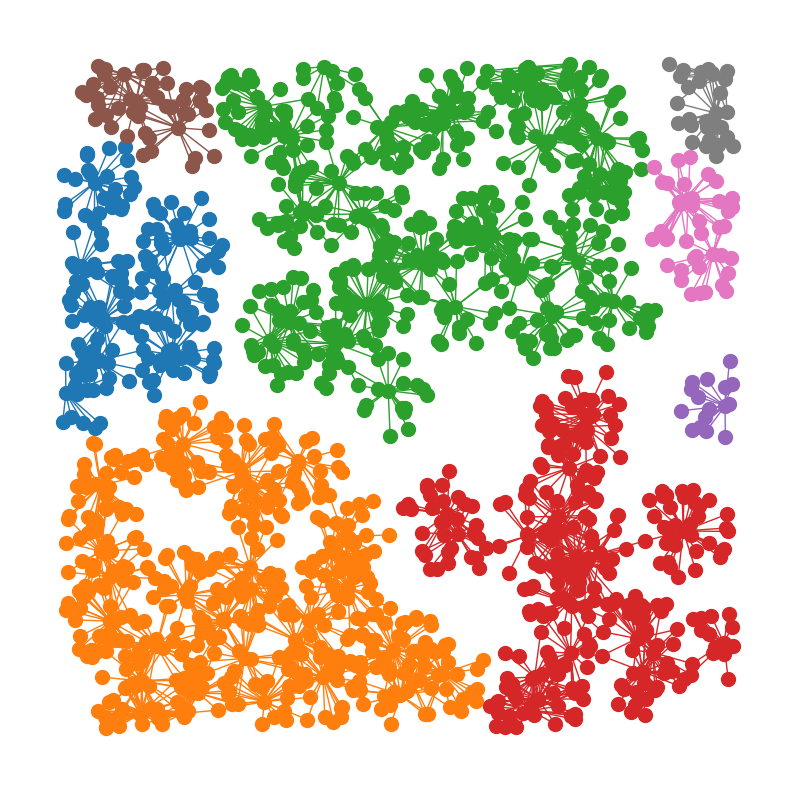} \\
\mbox{(a) $r = 0.05$} & \mbox{(b) $r = 0.55$} & \mbox{(c) $r = 0.85$} \\
\includegraphics[width=0.3\linewidth,height=1.0in]{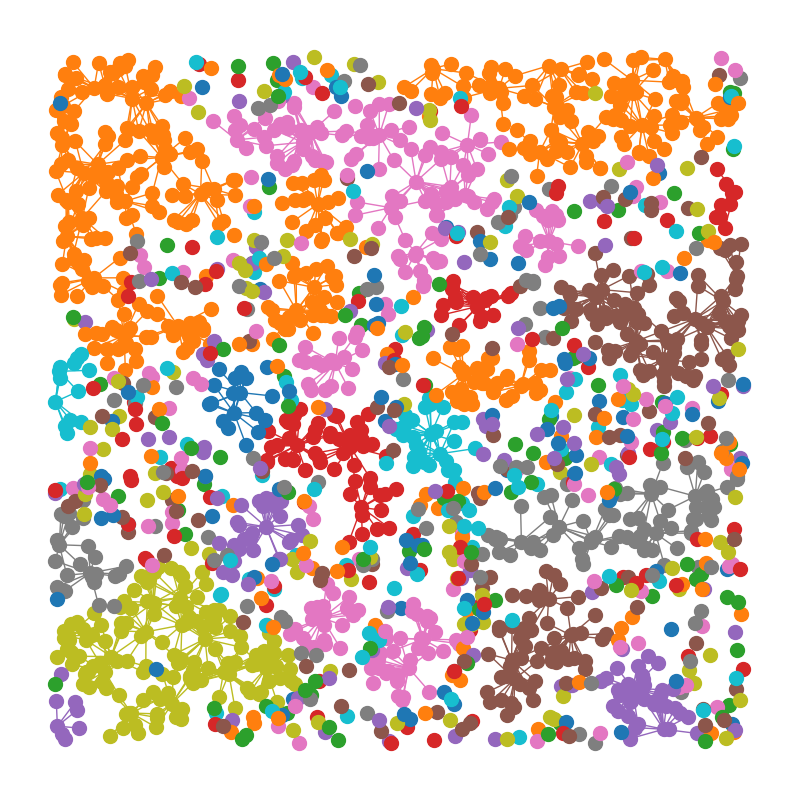}  &
\includegraphics[width=0.3\linewidth,height=1.0in]{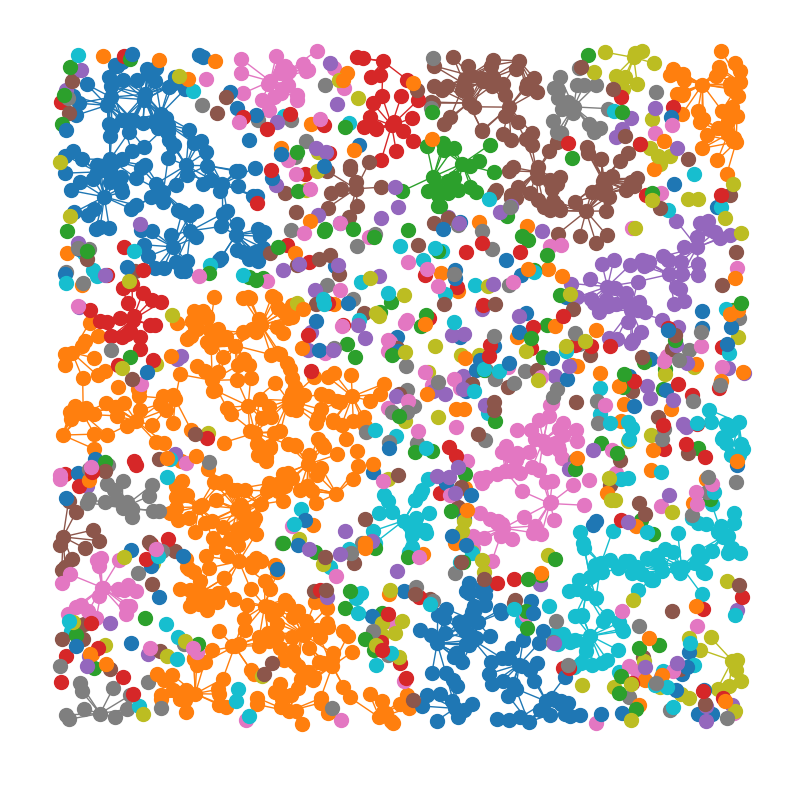}  &
\includegraphics[width=0.3\linewidth,height=1.0in]{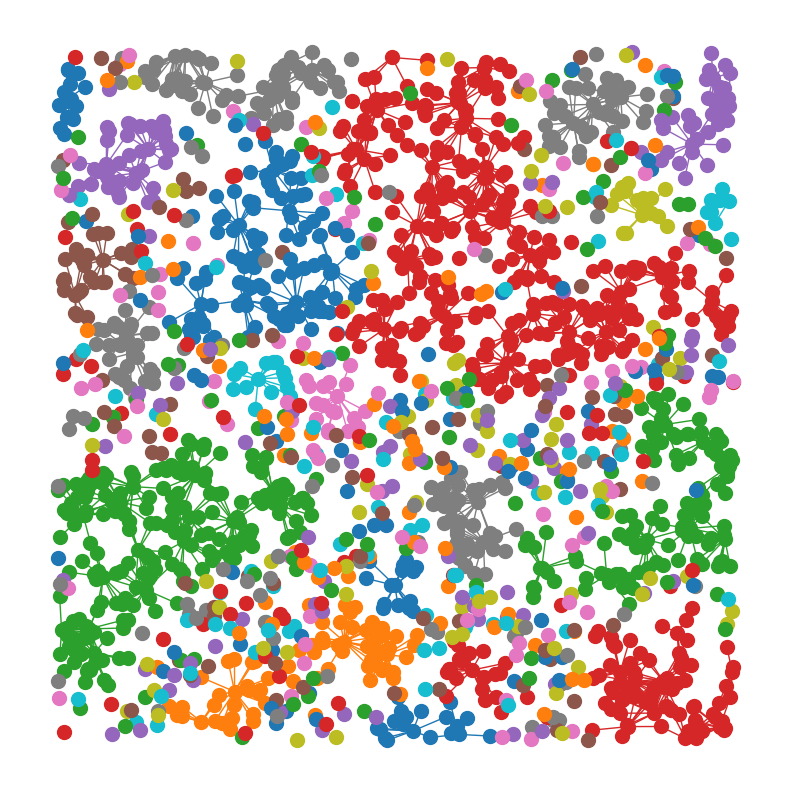} \\
\mbox{(a) $v = 0.3$} & \mbox{(b) $v = 0.7$} & \mbox{(c) $v = 1.1$} \\
\includegraphics[width=0.3\linewidth,height=1.0in]{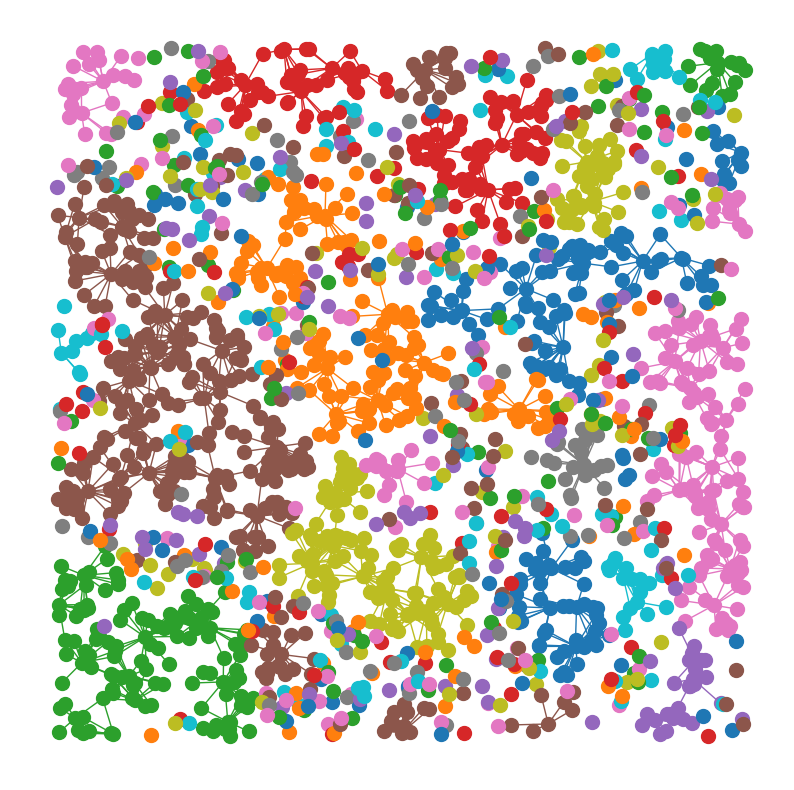}  &
\includegraphics[width=0.3\linewidth,height=1.0in]{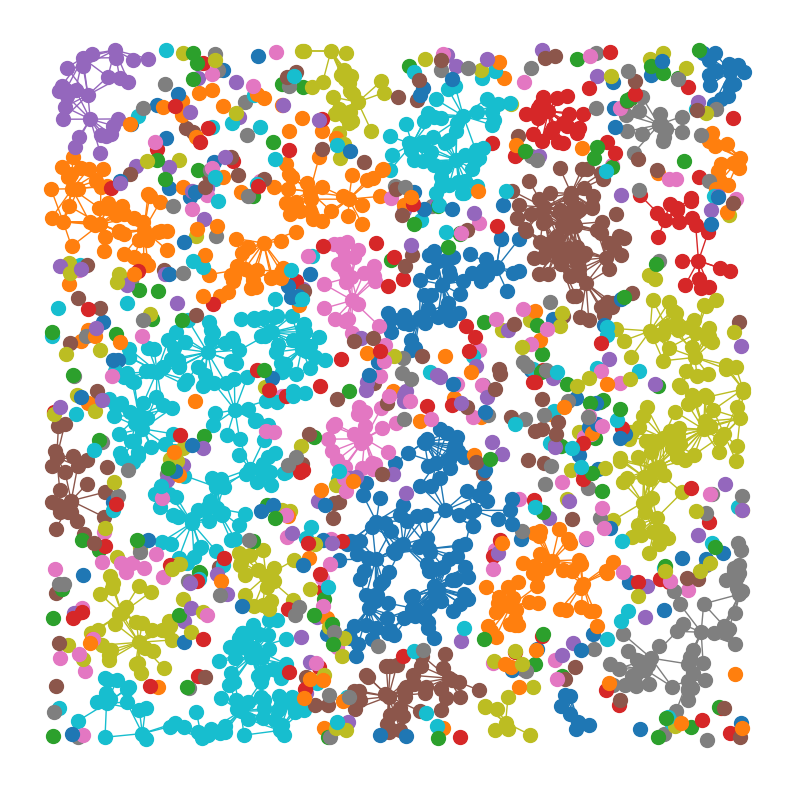}  &
\includegraphics[width=0.3\linewidth,height=1.0in]{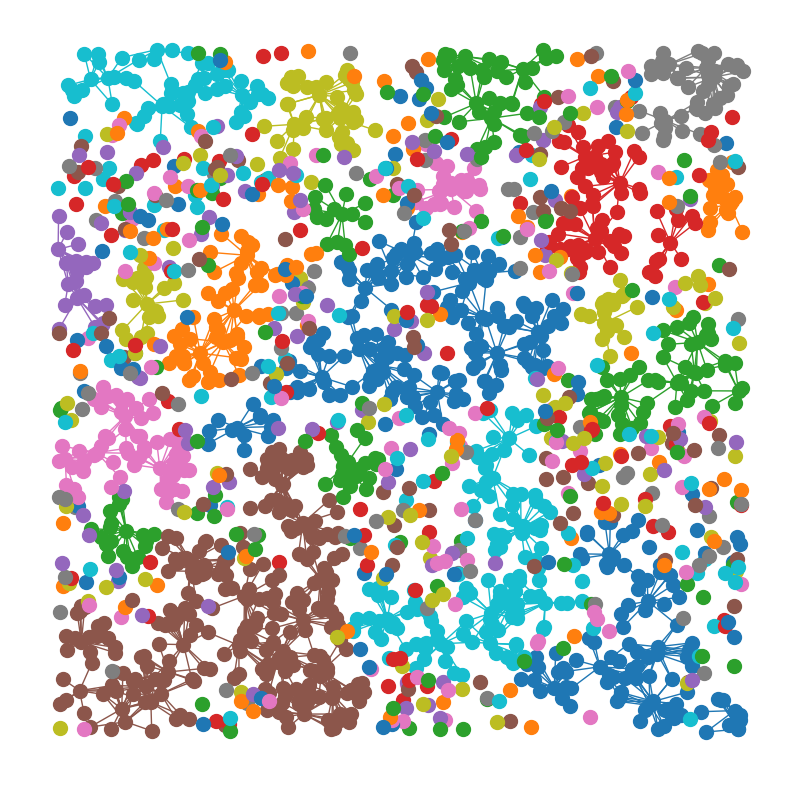} \\
\mbox{(a) $t_{rest} = 5$} & \mbox{(b) $t_{rest} = 15$} & \mbox{(c) $t_{rest} = 25$} \\
\end{array}$
\caption{Random geometric network having different number of connected components for varying radius, velocity and $t_{rest}$.}
\label{RGG_cluster}
\end{center}
\end{figure}

From Fig. \ref{RGG_cluster}, it can be analyzed that the total number of connected components depend on the values of radius, velocity and $t_{rest}$. With high $t_{rest}$ and low velocity, the network will be very stable, with nodes moving infrequently and only short distances. This results in a static or slowly evolving network with stable clusters. Whereas with high $t_{rest}$ and high velocity, nodes move infrequently but cover large distances when they do. This causes sudden and significant changes in the network structure, with stable periods interrupted by large reorganizations. Another scenario can be low $t_{rest}$ and low velocity where nodes move frequently but only short distances. This leads to a network that evolves gradually and continuously, with small changes accumulating over time. With low $t_{rest}$ and high velocity, the network will be highly dynamic, with nodes moving frequently and covering large distances. This creates a highly fragmented network with rapid changes in connections and clustering.

\section{Conclusions and Future Scopes}
In this paper, we proposed a model for Geometric Networks with Mobile Nodes (GNMN) and examines key network properties such as connectivity, degree distribution, second hop neighbors, and centrality measures. The study finds that network parameters like radius ($r$), velocity ($v$), and time of rest ($t_{rest}$) influence these properties. Specifically, an increase in $r$ makes the network denser, while higher $v$ and $t_{rest}$ result in frequent connections and disconnections without significantly affecting network density. The number of second hop neighbors shows similar trends. Increasing $r$ leads to a steady rise in connectivity and degree distribution, whereas changes in $v$ and $t_{rest}$ produce perturbed curve of degree distributions. Degree centrality scores rise exponentially with $r$ and linearly with $v$ and $t_{rest}$. Betweenness centrality fluctuates due to the dynamic appearance and disappearance of nodes in the communication among the nodes in the network. The network becomes more clustered with higher $r$, and it remains very stable with high $t_{rest}$ and low velocity.

In near future, we will apply epidemic spreading and information propagation on GNMN considering all the mentioned parameters and solve some real life problem such as wireless communication, sensor networks, and robotics.

\bibliographystyle{splncs04}
\bibliography{RGN_paper}

\end{document}